\documentclass[aps,prl,twocolumn,longbibliography,superscriptaddress]{revtex4-2}

\usepackage{amsmath,amssymb,amsthm,amsfonts,mathrsfs,dsfont,braket,array,float}
\usepackage{color,xcolor,graphicx}
\usepackage{slashed}
\usepackage{enumitem}
\usepackage{stmaryrd} 
\usepackage{esint}
\usepackage{physics}
\usepackage{braket}
\usepackage{tensor}
\usepackage{float}
\usepackage{mathtools}
\usepackage{bbold}
\usepackage{shortcuts}
\usepackage{microtype}   

\definecolor{rossocorsa}{rgb}{0.83, 0.0, 0.0}
\definecolor{navyblue}{rgb}{0.0, 0.0, 0.5}
\definecolor{custom}{rgb}{0.05,0.31,0.55}

\usepackage[
  bookmarks=true,
  bookmarksnumbered=false,
  hyperindex=true,
  bookmarksopen=true,
  hyperfigures=true,
  colorlinks=true,
  linkcolor=navyblue,
  citecolor=rossocorsa,
  urlcolor=custom,
  breaklinks]{hyperref}

\begin{document}

\title{Baby Universes from Thermal Pure States in SYK}

\author{Martin Sasieta}
\email{msasieta@berkeley.edu}
\affiliation{Leinweber Institute for Theoretical Physics, University of California, Berkeley, California 94720, USA}

\author{Brian Swingle}
\email{bswingle@brandeis.edu}
\affiliation{Martin Fisher School of Physics, Brandeis University, Waltham, Massachusetts 02453, USA}

\author{Alejandro Vilar López}
\email{alejandro.vilarlopez@ubc.ca}
\affiliation{Department of Physics and Astronomy, University of British Columbia, 6224 Agricultural Road, Vancouver, B.C. V6T 1Z1, Canada}

\begin{abstract}
We construct a simple two-dimensional holographic model of a closed ``baby'' universe. The baby universe spacetime originates from the black hole interior in Jackiw-Teitelboim (JT) gravity. The holographic description is a low-temperature thermal pure state of two Sachdev-Ye-Kitaev (SYK) models. The construction of the state relies on the insertion of a heavy matter operator, which supports the baby universe, together with a first-order thermal phase transition between a black hole phase and an empty Anti de Sitter (AdS) phase. For the transition, we employ a version of the Maldacena-Qi phase transition of two coupled SYK systems. The bulk entanglement between the AdS region and the baby universe remains $O(N)$ below the transition, owing to the large number of light bulk fields. This entanglement admits a definition through coarse-graining over SYK couplings.
\end{abstract}

\maketitle

\textbf{Introduction} — Formulating a consistent paradigm for quantum gravity in closed universes is one of the central challenges of holography, as our own Universe might well be closed. The main obstacle is the absence of a rigid spatial boundary on which to define the theory. Nevertheless, insights from the Anti–de Sitter/Conformal Field Theory (AdS/CFT) correspondence provide a useful guide. Among these is the observation that semiclassical baby universes with a negative cosmological constant are obtained by analytically continuing Euclidean AdS wormhole saddle points of the gravitational path integral \cite{Maldacena:2004rf,McInnes:2004ci,Cooper:2018cmb,Chen:2020tes,VanRaamsdonk:2020tlr,VanRaamsdonk:2021qgv,Antonini:2022blk}. These setups provide clean theoretical laboratories to formulate new proposals \cite{Abdalla:2025gzn,Harlow:2025pvj}.

The most puzzling aspect of these constructions is that the wormhole solutions preparing gravitational states of the closed universe arise from an (implicit or explicit) disorder average over products of partition functions or correlation functions in the CFT \cite{Saad:2018bqo,Saad:2019lba,Pollack:2020gfa,Belin:2020hea,Marolf:2020xie,Stanford:2020wkf,Chandra:2022bqq,Sasieta:2022ksu}. Before the average, each of these quantities corresponds to an ordinary $\mathbb{C}$-number. While it is unclear whether the wavefunction of the closed universe has a meaningful non-perturbative definition at the level of the microscopic numbers themselves, it clearly emerges through their collective statistics.

This picture departs sharply from the way the bulk emergence is understood in AdS/CFT. In \cite{Antonini:2023hdh} (hereafter AS$^2$), a simple model was proposed, making a direct connection between the emergence of baby universes and the supposedly more conventional holographic encoding of the black hole interior. The central insight is that the black hole interior is a cosmological spacetime, and, in particular, one that becomes spatially closed if the black hole fully evaporates. Instead of tackling the subtleties of dynamical evaporation, the AS$^2$ model employs the first-order Hawking–Page phase transition between the black hole phase and the particle gas phase. Starting from states that describe black holes with macroscopic interiors supported by a thin shell of matter, cooling them down across the transition produces a gravitational spacetime that appears to contain a closed universe as a result of the spatial disconnection of the black hole interior. These states are prepared by Euclidean wormhole-like saddle points of the gravitational path integral. See \cite{Antonini:2023hdh,Antonini:2025ioh} for details and \cite{Antonini:2024mci,Engelhardt:2025vsp,Higginbotham:2025dvf,Antonini:2025ioh,Liu:2025cml,Gesteau:2025obm,Kudler-Flam:2025cki,Mori:2025jej,Belin:2025ako} for recent literature involving AS$^2$.

In this Letter, we propose a simpler two-dimensional model of baby universes that retains many of the key features of the AS$^2$ construction. The model consists of a two-dimensional baby universe (the ``bulk'' system) holographically described by a non-gravitating quantum mechanical system (the ``microscopic'' or ``boundary'' system). Our aim is twofold. On the one hand, the model potentially allows us to study several subtleties that arise in AS$^2$, especially those related to the microscopic matrix elements of the heavy operator dual to the thin shell. On the other hand, the present model displays additional appealing properties and can be worked out in full detail. An extensive analysis of these features will be presented elsewhere \cite{usprep}.\\

{\it Note added:} As this manuscript was being finalized, we became aware of independent, related upcoming work on baby universe emission in black hole evaporation \cite{Ahmedtoappear}. \\

\textbf{Microscopic model} — We start with the Sachdev-Ye-Kitaev (SYK) model, a quantum system of $N$ Majorana fermions $\psi_1, \ldots, \psi_N$ satisfying $\{\psi_i, \psi_j\} = 2\delta_{ij}$. The fermions interact through a $q$-body Hamiltonian \cite{Kitaev:2015SYKTalks,Maldacena:2016hyu}
\be\label{eq:SYKH}
H_{\text{SYK}} = \iw^{\frac{q}{2}} \sum_{i_1<...<i_q} J_{i_1...i_q}\, \psi_{i_1}...\psi_{i_q}\,.
\ee
The couplings $J_{i_1...i_q}$ are independent Gaussian random variables with zero mean and variance
\be
\overline{J_{i_1...i_q} J_{i'_1...i'_q}} = \frac{\mathcal{J}^2 N}{2q^2{N \choose q}}\, \delta_{i_1,i'_1}...\delta_{i_q,i'_q}\,.
\ee
The infrared of the SYK model collectively describes the universal gravitational mode of near-extremal black holes with Schwarzian action.

To construct our model, we require $\textit{(i)}$ a heavy operator that provides the necessary backreaction to support the baby universe, and $\textit{(ii)}$ a thermal first-order phase transition between a gapless “black hole” phase and a gapped “empty AdS” phase.

For $\textit{(i)}$ we can take a Majorana string
\be\label{eq:Mstring} 
\Op_\Delta = \iw^{\frac{r}{2}}\psi_{i_1}...\psi_{i_r}
\ee 
which has a conformal dimension of $\Delta = r/q$ in the infrared. We may continue into the regime $r = O(N)$ so that the resulting operator is heavy, although this may entail corrections to the conformal dimension. More generally, for such long strings one should not expect a sharp identification with a single heavy infrared conformal primary. In this Letter, we will assume that its coarse-grained large-$N$ effect is nevertheless well captured by an effective heavy particle/worldline description, an assumption that appears to receive semi-quantitative support \cite{Bettaque:2026vpl}.

For $\textit{(ii)}$ we consider two SYK models and define the Maldacena–Qi (MQ) Hamiltonian \cite{Maldacena:2018lmt} 
\be\label{eq:MQH}
H = H_{\text{SYK},\mathsf{L}} + H_{\text{SYK},\mathsf{R}}^*
	-	\frac{\mu N}{{N \choose p}}\, (-\iw)^{p^2}\!\sum_{i_1<...<i_p}\!\!
\psi^{\mathsf{L}}_{i_1...i_p}\psi^{\mathsf{R}}_{i_1...i_p} \,,
\ee
where we use the shorthand notation $\psi_{i_1...i_p} \equiv \psi_{i_1}...\psi_{i_p}$.

The coupled SYK models interacting via \eqref{eq:MQH} undergo a first order thermal phase transition at the coupling-dependent MQ inverse temperature $ \mathcal{J}\beta_{\ast}$. In the low-temperature phase, the ground state $\ket{\rm 0}$ and gapped excitations dominate, and these correspond to a connected semiclassical wormhole along with its matter and gravitational excitations. This characterization is valid as long as the interaction strength is small, but not too small \cite{Maldacena:2018lmt}
\be\label{eq:regime}
\dfrac{1}{N^{2(1-\Delta_H)}}\ll \dfrac{\mu}{\mathcal{J}} \ll 1\,\qquad N\to \infty\,,
\ee 
where $\Delta_H \equiv p/q$ is the infrared conformal dimension of the Majorana string $\psi_{i_1...i_p}$ in the MQ interaction \eqref{eq:MQH}.

\vspace{.2cm}

\textbf{Thermal pure states} — In this setup, we define a \emph{thermal pure state} (TPS) of the MQ Hamiltonian as
\be\label{eq:TPS}
\ket{\Psi^\Delta_\beta} \propto e^{-\frac{\beta}{2}H}\ket{\Op_{\Delta}}
= \sum_n e^{-\frac{\beta}{2}E_n} c_n \ket{E_n}\,,
\ee
where $\ket{E_n}$ are the energy eigenstates of $H$. { The reference state $\ket{\Op_\Delta} \equiv \Op_\Delta^\mathsf{L}\ket{I}$ is the (unnormalized) two-sided state associated with the single-sided operator $\Op_\Delta$ by the operator-state isomorphism. It is obtained by acting with $\Op_\Delta$ on the left SYK system of the (unnormalized) infinite-temperature thermofield double state $\ket{I}$ of the two SYK models.} The latter is defined as the $\mathsf{LR}$ Dirac fermion ground state
$(\psi_i^{\mathsf{L}} - \iw\,\psi_i^{\mathsf{R}})\ket{I} = 0$,  $\;\forall\;i = 1,\ldots,N$.  The coefficients in \eqref{eq:TPS} are therefore
\be
c_n = \braket{E_n}{\Op_{\Delta}}\,.
\ee
The canonical thermal state of the MQ Hamiltonian is recovered by averaging the TPS over all choices of $\mathcal{O}_\Delta$.

The Euclidean preparation of the TPS is schematically represented as\\[-.7cm]
\begin{center}
\includegraphics[width=0.5\linewidth]{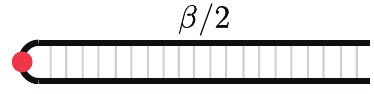}
\end{center}
where the thick lines indicate Euclidean time in the two SYK systems, the thin lines represent the MQ coupling, and the red dot denotes $\mathcal{O}_\Delta$.

In the absence of MQ coupling ($\mu = 0$), the TPS reduces to the \emph{partially entangled thermal state} (PETS) of two SYK models, introduced in \cite{Goel:2018ubv} (see \cite{Balasubramanian:2022gmo,Balasubramanian:2022lnw} for higher-dimensional versions). The TPS defined here is distinct from variants based on Haar random states~\cite{PhysRevA.49.668,Sugiura_2013}.

To understand the semiclassical state dual to the TPS, we examine the dominant support of its wavefunction in the large-$N$ limit. This is determined by the norm of the state, given by the survival amplitude
\be 
\mathscr{N}^{\Delta}_\beta = \bra{\Op_\Delta} e^{-\beta H}\ket{\Op_\Delta} = \sum_{n} e^{-\beta E_n} |c_n|^2\,.
\ee 

Much like the canonical partition function (the case where $|c_n|^2 =1$), in the large-$N$ limit, $ \mathscr{N}^{\Delta}_\beta$ exhibits a first order phase transition as a function of $\beta$ at the MQ inverse temperature $\beta_{\ast} = O(N^0)$. The details of the transition here depend on the operator, which we consider to be heavy enough to modify the large-$N$ saddle points. 

Above the transition temperature $\beta< \beta_{\ast}$ (still $\beta \mathcal{J}\gg 1$) the wavefunction of the TPS is dominated by a dense spectrum. Below the transition, $\beta> \beta_{\ast}$, it is dominated by the ground state and gapped quasi-particle excitations
\be
e^{\beta E_0}\mathscr{N}^\Delta_\beta = |c_0|^2 + e^{-\beta E_{\rm gap}} \sum_{n=1}^{N_*} |c_n|^2 + ... \,,
\ee 
where $c_0 = \bra{\rm 0}\ket{\Op_{\Delta}}$ is the ground state amplitude. Here $E_0$ is the ground state energy, $E_{\rm gap} =O(N^0)$ is the gap, and $N_* = N$ is the number of first excited states (fundamental excitations) of the MQ Hamiltonian. The dots represent contributions of multi-particle excitations. 

We next discuss the corresponding bulk description, which is expected to describe a suitably averaged (over couplings, heavy operators, etc.) form of this microscopic norm. We discuss the physical meaning of the average after presenting the semiclassical bulk description.\\

\textbf{Semiclassical state} —  The bulk theory is Jackiw-Teitelboim (JT) gravity with matter, $I = I_{\text{JT}}+ I_{\text{matter}}$. The JT part of the action,
\be\label{eq:JTaction}
I_{\text{JT}} = -S_0 \,\chi[\Sigma] - \frac{1}{2}\!\int_\Sigma\! \Phi(R+2) - \!\int_{\partial \Sigma}\!\Phi_b (K-1),
\ee
describes the topological and boundary Schwarzian contributions, with $\chi[\Sigma]=2-2g-n$ the Euler characteristic of the manifold $\Sigma$ of genus $g$ and $n$ boundaries. The gravitational coupling to matter is set by the boundary value of the dilaton, $\Phi_b$. The relation of this parameter to SYK scales is 
\be 
\Phi_b = \dfrac{N\alpha_S}{\mathcal{J}}\,,
\ee
where $\alpha_S$ is a $q$-dependent constant that can be found in \cite{Maldacena:2016hyu}. We tune the topological factor $S_0$ to the large-$N$ ground-state entropy of SYK \cite{Maldacena:2016hyu}. For the matter sector, we consider minimally coupled but otherwise general matter with action $I_{\rm matter}$; { to make more detailed contact with the microscopic model, we will assume that it is effectively described by $O(N)$ bulk massive fields, although the precise bulk dual of SYK is not fully understood.}

In the bulk, the gravitational path integral evaluation of the coupling-averaged norm $\overline{\mathscr{N}^\Delta_\beta}$, via saddle point evaluation, computes the norm of the semiclassical state dual to the TPS. We will find two relevant competing saddle point contributions as $N\rightarrow \infty$,
\be 
\overline{\mathscr{N}^\Delta_\beta} \sim e^{-I_\beta[\Sigma_{\rm d}]} Z_{\rm d} + e^{-I_\beta[\Sigma_{\rm h}]} Z_{\rm h}\,.
\ee 
The first saddle $\Sigma_{\rm d}$ has disk topology ($g=0,n=1$) while the second saddle $\Sigma_{\rm h}$ has the topology of a disk with a handle ($g=1,n=1$). Here $Z_{\rm d},Z_{\rm h}$ are bulk partition functions on these manifolds. The difference in classical actions $I_\beta[\Sigma_{\rm h}] - I_\beta[\Sigma_{\rm d}]$ has an extensive positive ``entropic'' contribution from the topological factor $2S_0$, but it receives an extensive negative ``energetic'' contribution from the MQ coupling, of order $\beta E_{0}$ (recall $|E_{0}| = O(N)$). Roughly, the transition between the two saddle points occurs at an inverse temperature $\beta_{\ast} \sim 2S_0/|E_0| = O(N^0)$.\\

\begin{figure}[h]
    \centering
    \includegraphics[width=0.95\linewidth]{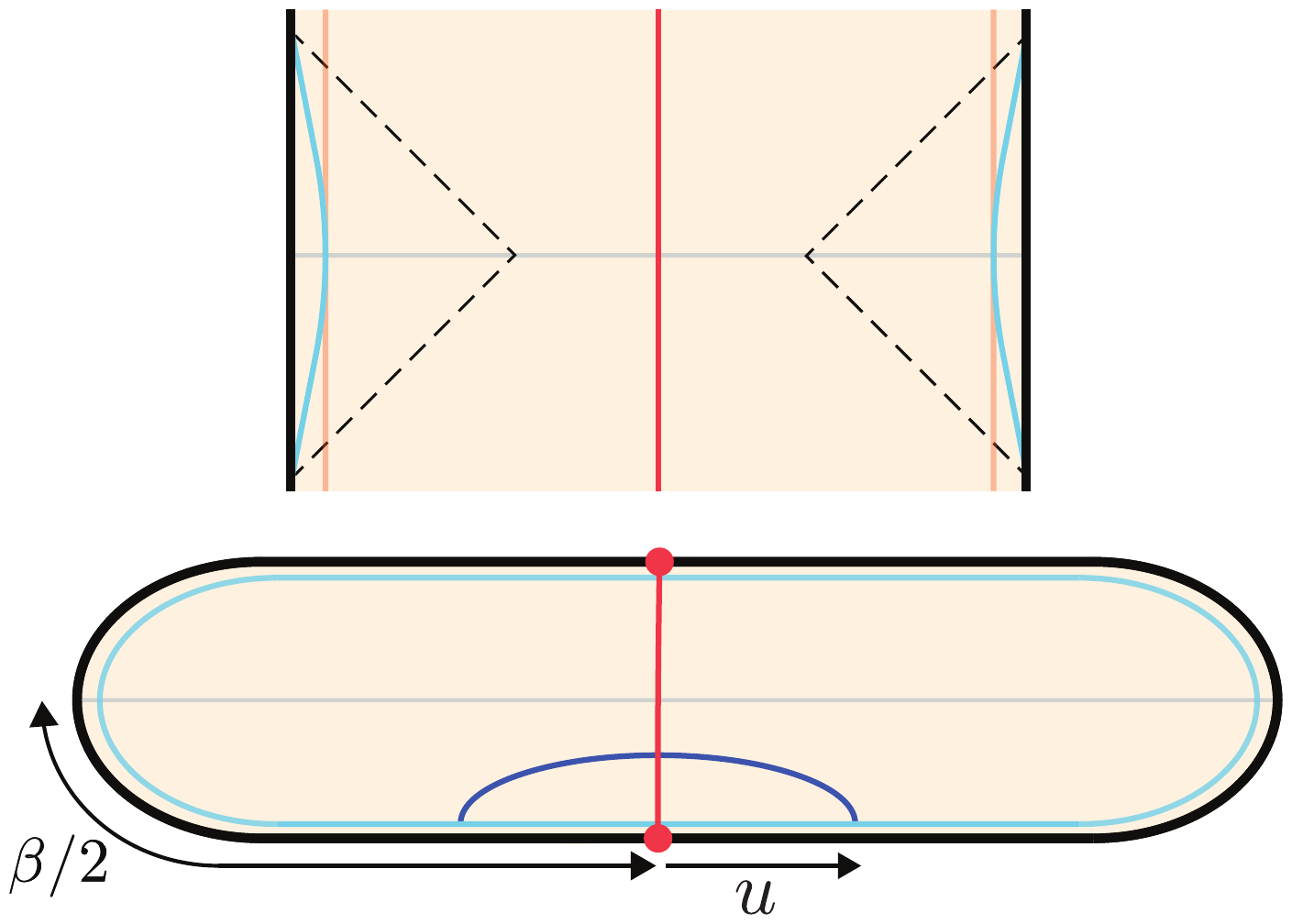}
    \caption{\textbf{Top:} Semiclassical state of a two-sided black hole with a heavy matter particle in the black hole interior. Evolving the state with a strong enough MQ coupling (orange) allows us to reconstruct the interior. \textbf{Bottom:} The Euclidean manifold $\Sigma_{\rm d}$ preparing the semiclassical state on the gray slice. The geodesic of renormalized length $\ell$ is shown in dark blue.}
    \label{fig:petssaddle}
\end{figure}

{\it Black hole saddle.} We first consider the saddle $\Sigma_{\rm d}$ of disk topology, sketched in Fig. \ref{fig:petssaddle}. This saddle prepares the gravitational state of a two-sided black hole on the time-symmetric (gray) slice. One could neglect the MQ interaction to study this saddle using the analysis of \cite{Goel:2018ubv}. Here, we outline a more precise treatment.

We consider the renormalized geodesic length $\ell(u)$ between two points to the left and right of the heavy insertion, evolved by boundary time $u_{\mathsf{L}} = u_{\mathsf{R}} =u$. From now on, we measure all dimensionful quantities, such as $u$, in terms of the SYK time $\alpha_S/\mathcal{J}$. The heavy operator inserts a lowest-weight state of the $\mathsf{PSL}(2,\mathbb{R})$ isometries, with a weight of $\Delta$. Enforcing gauge invariance through the vanishing of the total charges, one finds the effective action \cite{Maldacena:2018lmt}
\be 
I = \dfrac{N}{2} \int \text{d}u  \left(\dfrac{1}{2}\dot{\ell}^2 + V(\ell)\right) - S_0   \,,
\ee 
for the potential
\be\label{eq:potential} 
V(\ell) = 2e^{-\ell} - {\eta}  e^{-\Delta_H \ell } + \frac{2\Delta}{N} e^{-\ell/2}\,.
\ee 
Here $\eta \,=\, { \xi\mu}/{\mathcal{J}}$ is the dimensionless interaction strength, where $\xi$ is a constant that depends on $\alpha_S$ and can be found in \cite{Maldacena:2018lmt} (we are using the conventions of \cite{Magan:2024aet}).

The semiclassical solution is a bounce of the $\ell$-particle in the Euclidean potential $-V(\ell)$. We include the details of this solution in the Appendix. The effect of the heavy operator is that it contributes positively to the potential \eqref{eq:potential}. This causes the value of $\ell$ at the turning point to increase, and thus stretches the wormhole. As represented in Fig. \ref{fig:petssaddle}, the Lorentzian analytic continuation of the solution includes a long spatial wormhole at the time-symmetric slice. 

The classical action of this saddle has a topological term $-S_0$ and a contribution from the $\ell$-bounce. When the effects of the heavy operator dominate, this contribution should approach the action of a black hole saddle with a matter insertion. In the Appendix, we present the explicit solution for $\Delta_H = 1/2$, which results in an action
\be\label{eq:actiond} 
I[\Sigma_{\rm d}] = - S_0 - \frac{4 \pi^2 N}{\beta} + 2 N \kappa \left( 1 - \log \frac{N \kappa}{4} \right) \, ,
\ee 
where $\kappa = \Delta / N - \eta/2 > 0$ and we have expanded at a low enough preparation temperature, $\beta \kappa \gg 1$.\\

\begin{figure}[h]
    \centering
    \includegraphics[width=0.93\linewidth]{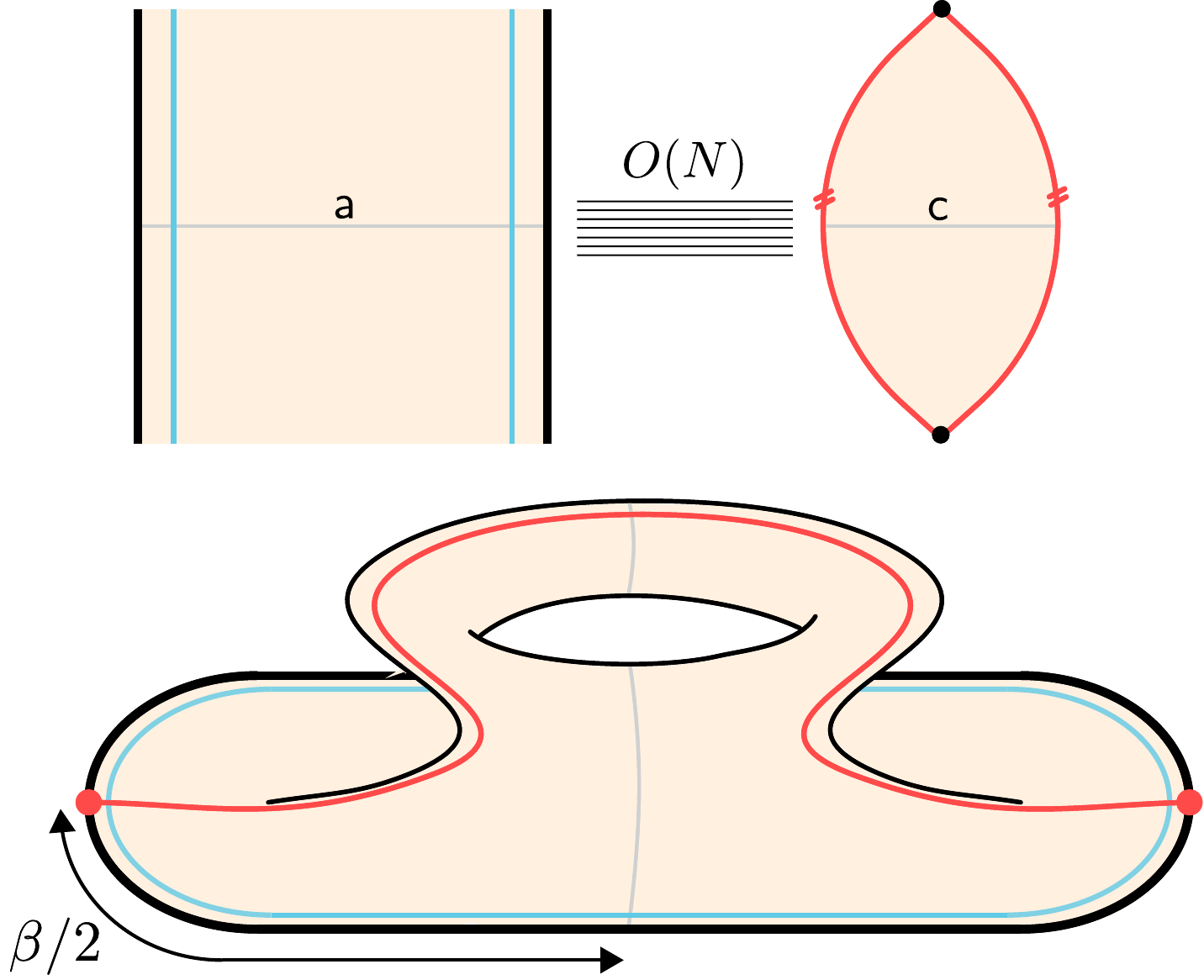}
    \caption{\textbf{Top:} Semiclassical state of an empty wormhole $\mathsf{a}$ with large bulk entanglement to a baby universe $\mathsf{c}$. \textbf{Bottom:} The Euclidean manifold  $\Sigma_{\rm h}$ preparing the semiclassical state on the gray slice. The moduli of the handle are stabilized by the heavy particle and by the MQ interaction. }
    \label{fig:cosmosaddle}
\end{figure}

{\it Baby universe saddle.} We now consider the saddle $\Sigma_{\rm h}$ of disk topology with a handle, as sketched in Fig. \ref{fig:cosmosaddle}. This saddle prepares a gravitational state on empty AdS$_2$ ($\mathsf{a}$) and a baby universe ($\mathsf{c}$). { The handle is stabilized by both the MQ interaction and the heavy operator: the MQ coupling fixes the separation between the two mouths of the wormhole, while the heavy operator fixes the bottleneck modulus. Without the MQ coupling, or if the particle does not propagate through the wormhole, this geometry would not be on shell and its SYK interpretation would be unclear. In the present setup, it gives the contribution of the low-temperature ``empty wormhole'' phase to the TPS wavefunction \eqref{eq:TPS}.}

The details of this saddle are presented in the Appendix. Essentially, for large $\beta$, the handle can be decomposed into two pieces. One piece is a bounce of the $\ell$-particle in the potential \eqref{eq:potential} with $\Delta=0$. This new potential has a minimum associated with the semiclassical ground state of the MQ Hamiltonian. The $\ell$-particle spends most of the Euclidean time in this configuration. The MQ energy of this configuration is
\be\label{eq:energy} 
 E_0 = -N \tfrac{1-\Delta_H}{\Delta_H}(\tfrac{1}{2}\eta \Delta_H)^{\frac{1}{1-\Delta_H}}\,.
\ee 

For large $\beta$ compared with the inverse gap of the MQ Hamiltonian, the previous piece of the handle prepares the MQ ground state. Since this is close to a finite-temperature TFD state of two decoupled SYK systems, at the classical level, the second piece corresponds to the wormhole arising in the squared thermal one-point function of the operator \cite{Stanford:2020wkf,Usatyuk:2024mzs}. { The corresponding effective inverse temperature $\beta_0$ is fixed by matching the average energy of the MQ ground state, as measured by the decoupled SYK Hamiltonians, to that of the TFD. The effective inverse temperature of the thermal one-point function is set by \cite{Maldacena:2018lmt}}
\be\label{eq:invtemp} 
\beta_0 = 2\pi \sqrt{\dfrac{N}{|E_0|}\dfrac{1-\Delta_H}{\Delta_H}}\,.
\ee
 
We denote by $b(\Delta,\eta)$ the geometric modulus in $\Sigma_{\rm h}$ that determines the size of the baby universe component $\mathsf{c}$. This modulus is the length of the bottleneck of the wormhole (closed gray geodesic in Fig. \ref{fig:cosmosaddle}). In AdS units, this length scale is fixed by the mass of the operator and by the MQ coupling (in the regime $\Delta \beta_0/N\gg 1$)
\be\label{eq:sizebu} 
b(\Delta,\eta) \approx 2 \log \dfrac{\Delta}{N} - \dfrac{1}{1-\Delta_H} \log  \frac{\eta \Delta_H}{2}  \,.
\ee
Upon Lorentzian evolution, this part of the geometry becomes a crunching cosmology, as illustrated in Fig.~\ref{fig:cosmosaddle}.

The total action of the saddle receives a contribution $S_0 + \beta E_0$ from the topological piece and the Euclidean evolution close to the ground state of the MQ Hamiltonian. To compare with \eqref{eq:actiond}, for $\Delta_H = 1/2$ we have
\be\label{eq:actionh}
I[\Sigma_{\rm h}] \approx S_0 + \beta E_0 + 2 N \kappa \left( 1 - \log \frac{N \kappa}{4} \right) \, .
\ee

{\it Phase transition.} From \eqref{eq:actiond} and \eqref{eq:actionh} the large-$N$ action difference $I_\beta[\Sigma_{\rm h}] - I_\beta[\Sigma_{\rm d}] $ changes sign for $\beta_\ast = O(N^0)$. If we neglect all the terms except for the entropy $2S_0$ and the MQ ground state energy $\beta E_0$, we obtain the transition temperature $\beta_\ast \approx 2S_0/|E_0|$ of the MQ partition function. Including the rest of the terms yields
\be 
\beta_\ast \approx \dfrac{S_0}{|E_0|}\left(1 + \sqrt{1+\dfrac{4\pi^2 N |E_0|}{S_0^2}}\right)\,.
\ee 
Note that in this approximation, where $\Delta \beta_0/N \gg 1$,  the transition becomes independent of $\Delta$.\\

\textbf{Bulk state \& holographic map} — Even though the microstate $|\Psi^\Delta_\beta\rangle$ is pure, for $\beta> \beta_\ast$, the state of the bulk quantum fields $\ket{\psi_\beta} \in \mathcal{H}_{\mathsf{a}} \otimes \mathcal{H}_{\mathsf{c}} $ appears highly entangled with the baby universe $\mathsf{c}$, exhibiting $O(N)$ entanglement just below the transition temperature. Here, the extensivity of the entanglement entropy follows from the $O(N)$ number of light matter species in the bulk.

More precisely, the state is prepared by the path integral of the bulk fields in half of $\Sigma_{\rm h}$. The state has the Schmidt form
\be\label{eq:TFDlikebulk} 
\ket{\psi_\beta}
= \frac{\Pi_q}{\sqrt{Z(\beta,q)}}
\sum_{n} e^{-\frac{\beta}{2}E_n}
\ket{E_n}_{\mathsf{a}}\ket{\varphi_n}_{\mathsf{c}}\,,
\ee 
where $\ket{\varphi_n}$ are states in the closed universe prepared by the wormhole specified in the Appendix (for sufficiently large $\Delta$, these are essentially copies of $\ket{E_n}$). Here, $\Pi_{q}$ is an orthogonal projector into a sector of fixed gauge charge (of the boundary global symmetry defined in the next section), and $Z(\beta,q) ={\rm Tr}\, (e^{-\beta H}\Pi_{q})$ is the canonical partition function of the MQ Hamiltonian within the sector. In particular, the reduced state $\rho_{\mathsf{a}} = \text{Tr}_{\mathsf{c}} \ket{\psi_\beta} \bra{\psi_\beta}\, \propto\, \Pi_q\rho_{\beta}\Pi_q$ is approximately thermal within the fixed-charge sector.  

Knowing both the bulk state and the microstate, we can identify the holographic map $W$ that relates them through $W\ket{\psi_\beta} = |\Psi^\Delta_\beta\rangle$. For the AdS$_2$ factor $\mathsf{a}$, the holographic isometry is simple: the gapped excitations in the bulk $\ket{E_n}_{\mathsf{a}}$ map to energy eigenstates of the MQ Hamiltonian in the gapped phase of the two SYK systems $\ket{E_n}$. Therefore, the holographic map consists of a projection onto the baby universe state
\be\label{eq:closeduniversestate}
\bra{\psi_{\Op}}: \mathcal{H}_{\mathsf{c}} \rightarrow \mathbb{C}\,,\qquad \bra{\psi_{\Op}}\ket{\varphi_n} = c_n \,.
\ee
The full holographic map is $W = \mathbf{1}\otimes \bra{\psi_{\Op}}$, where the identity factor must be understood in the gapped phase, for $E_n - E_0 \ll |E_0|$.

{ This microscopic projection onto a single state expresses the intrinsic ``one-dimensionality'' of the baby universe as described from the exterior holographic SYK Hilbert space. In particular, in the $\beta\to\infty$ limit, different semiclassical configurations of the closed universe, for example, those obtained by inserting light operators in the Euclidean past, are all mapped by the projection $\bra{\psi_{\Op}}$ to the same underlying SYK microstate, namely the MQ ground state. This is the sense in which the Gram-matrix rank calculations in the literature diagnose a one-dimensional Hilbert space. At finite $\beta$, by contrast, bulk entanglement permits a non-trivial holographic description of the closed universe. The Hilbert space dimension of this description is set by the microcanonical entropy of the MQ Hamiltonian, which remains large, $O(N)$, provided $\beta$ is not too large. Microscopically, this happens because, at finite $\beta$, inserting light matter operators in the Euclidean preparation generally produces different states from the original TPS, but with approximately the same energy.

An intrinsic Hilbert space ``grows'' once one allows for coarse-graining over microscopic data such as the couplings or the choice of heavy operator. In that case, the closed universe is described by, or lies in the entanglement wedge of, an auxiliary microscopic reference system that keeps track of this data (sometimes referred to as ``the observer''). As we explain below, the present model provides a concrete microscopic realization of this idea through the disordered couplings of the SYK model.}\\

\textbf{Discrete global symmetries} — The MQ Hamiltonian (for concreteness $q=4$ and $p=1$) possesses a discrete global $\mathbb{Z}_4$ symmetry associated with fermion number $Q = N - \iw \sum_{i=1}^N \psi_i^{\mathsf{L}}\psi_i^{\mathsf{R}}$ (mod $4$) \cite{Garcia-Garcia:2019poj}. If the exact correlation of the couplings of the $\mathsf{L}$ and $\mathsf{R}$ SYK models is absent, this symmetry reduces to fermion parity $(-1)^Q$.

As $\Op_\Delta$ is a Majorana string of a definite length $r$, such an operator will have a definite fermion number $r$, and so will the TPS \eqref{eq:TPS}. In this case, the superselection rule implies that the coefficients $c_n$ of the TPS wavefunction with a different charge identically vanish. In particular, since $\ket{0}$ has a vanishing global charge in the large-$N$ limit, the ground state component will vanish unless the string has a length that is a multiple of $4$. Likewise, if $Q\ket{E_n} = q_n \ket{E_n}$, we have
\be\label{eq:Z4gauge}
c_n= 0\,\Leftrightarrow\, q_n + r \not\equiv 0\;\; ({\rm mod\,}4)\,.
\ee 

According to the standard holographic dictionary, the semiclassical bulk theory describing the gapped phase will necessarily contain a discrete $\mathbb{Z}_4$ gauge field, with an associated Wilson line in any of the four possible representations. The handle cannot serve to propagate a net amount of gauge charge, since that would fail to satisfy Gauss' law. Therefore, the semiclassical wormhole solution will provide a vanishing contribution for bulk states in $\mathsf{a}$ satisfying \eqref{eq:Z4gauge} and $\Pi_{q} = \delta_{Q,q}$ with $q \equiv -r$ (mod $4$). It is interesting that this information about the baby universe can be accessed quite directly by observing that the corresponding TPS has a vanishing wavefunction in certain charge sectors. \\

\textbf{Baby universe wavefunction} — This model allows for a more microscopic study of the baby universe. In what follows, we fix a heavy Majorana string $\Op_\Delta$ \eqref{eq:Mstring} and numerically probe the resulting baby universe wavefunction $\ket{\psi_{\Op}}$, or equivalently, the wavefunction of the TPS $|\Psi_{\beta}^\Delta\rangle$, as a function of the SYK couplings for finite $N$ SYK. For concreteness, we focus on its ground state component,
\be
c_0[J] = \bra{\psi_{\Op};J}\ket{\varphi_0}\,,
\ee 
where $[J]$ denotes the collection of SYK couplings, and $\ket{\psi_{\Op};J}$ makes explicit the corresponding dependence of the baby universe state on the microscopic couplings.

\begin{figure}[h]
    \centering
    \includegraphics[width=\linewidth]{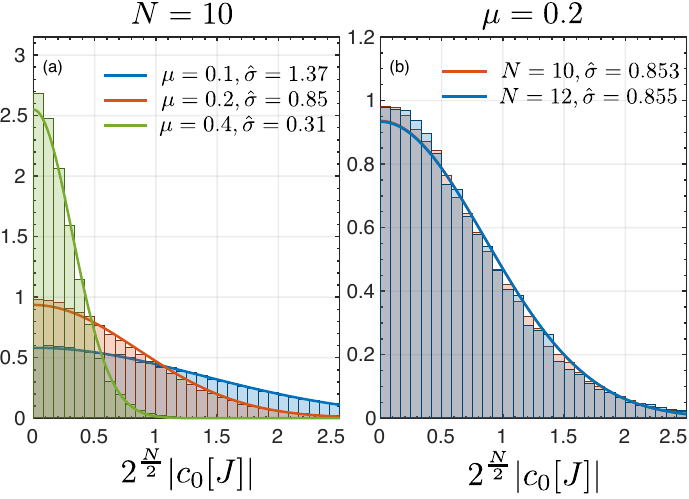}
    \caption{Probability distribution of $c_0[J]$ for a fixed operator 
$\Op_\Delta=\psi_{1}\cdots\psi_{8}$ over $10^5$ realizations of the SYK 
couplings. We take $\mathcal{J}=1$. The global phase of the MQ ground state 
$\ket{0}$ is fixed by the condition $\langle 0|I\rangle>0$, and the analysis 
is restricted to the $Q\equiv 0~(\mathrm{mod}\,4)$ subspace. 
In (a) we fix $N$ and vary $\mu$, while in (b) we fix $\mu$ and vary $N$. We show only the absolute value; we find that the sign is uniformly distributed.}
    \label{fig:c0}
\end{figure} 

We plot our results in Fig.~\ref{fig:c0}. We find that $c_0[J]$ is non self-averaging, so that the amplitudes of the normalized TPS vary erratically with the choice of couplings. It is well described by a real Gaussian random variable with zero mean and variance of order $2^{-N}$. Its reality reflects the fact that the MQ ground state $\ket{0}$ is a finite-temperature thermofield-double state (for a Hamiltonian that is a perturbation of $H_{\text{SYK}}$), so $c_0[J]$ is a thermal one-point function of $\Op_\Delta$. As $\mu$ is increased, the overlap approaches its vanishing infinite-temperature value and therefore decreases in magnitude.

{ We focus on $c_0[J]$ as a representative example. In our numerics, the higher coefficients $c_n[J]$ display similar erratic behavior, although a more detailed analysis of their full distributions goes beyond the scope of this Letter.}

 This erratic form of the wavefunction component $c_0[J]$ can be attributed to a ``heavy operator randomness hypothesis'', as has been conjectured in AS$^2$ for the matrix elements of the heavy operator \cite{Sasieta:2022ksu}. To study the operator dependence, one may alternatively fix a single disorder realization and perform a similar statistical analysis by varying the choice of Majorana strings of fixed length in the definition of $\Op_\Delta$. Here, the numerical evidence supports the hypothesis that the baby universe state, or the TPS, are quasi-random. In forthcoming work, the erratic nature of such one-point functions will be demonstrated microscopically at large $N$ in SYK~\cite{Bettaque:2026vpl}. \\

\textbf{Entanglement to the baby universe} — We now propose a definition of the entanglement between $\mathsf{a}$ and $\mathsf{c}$ microscopically in this model. It is important to note that this entanglement does not correspond to microscopic boundary entanglement, since the boundary state is pure.

This entanglement needs to be interpreted more as a coarse-grained entropy. In \cite{Antonini:2023hdh,Antonini:2025ioh}, this was defined from the perspective of coarse-graining over all the matrix elements of the heavy operator (see \cite{Sasieta:2022ksu}). However, the ensemble of heavy operators is rather implicitly defined from the bulk, so we would like a more controlled definition. In \cite{Kudler-Flam:2025cki}, the idea of averaging over different values of $N$ was explored but found to be unsuccessful, essentially because this would produce $O(1/N)$ variances. 

Here we briefly lay out another option that this model opens up, namely using the ensemble of SYK couplings to define this entanglement \footnote{We thank Douglas Stanford and Henry Lin for valuable suggestions on this point} (see \cite{Almheiri:2021jwq,Qi:2021oni} for related work). Defining the state $\rho[J] \equiv |\Psi_{\beta}^\Delta[J]\rangle \langle\Psi_{\beta}^\Delta[J]|$ in SYK ($\mathsf{sys}$), we take $K$ independent instances of the microscopic SYK couplings and attach a reference system ($\mathsf{ref}$) that keeps track of the couplings in a ``pointer basis''
\be 
\rho_K(\mathsf{sys \cup ref}) \equiv \dfrac{1}{K} \sum_{\alpha=1}^K \rho[J_\alpha]\otimes (\ket{J_\alpha}\bra{J_\alpha})_{\mathsf{ref}}\,.
\ee 
The reduced density matrix to $\mathsf{sys}$ is now mixed $\rho^{\mathsf{sys}}_K  \equiv\text{Tr}_{\mathsf{ref}} (\rho_K(\mathsf{sys \cup ref})) =   \tfrac{1}{K}\sum_{\alpha=1}^K \rho[J_\alpha]$. The entropy of this density matrix when $K\rightarrow \infty$ is the ``annealed entropy'' over the microscopic ensemble of TPSs. The proposal is that this reproduces the bulk entanglement
\be\label{eq:annealede}
\lim_{K\to\infty} S(\rho_K^{\mathsf{sys}})
{\,\approx\,}
S(\rho_{\mathsf{a}})\,,
\ee
where $S(\rho) = -\text{Tr}(\rho \log \rho)$ is the von Neumann entropy of $\rho$. We will leave a more detailed examination of this proposal to \cite{usprep}. For sufficiently large $K$, provided \eqref{eq:annealede} is satisfied, this means that the closed universe lies within the entanglement wedge of the reference system. \\

\textbf{Discussion} — The model presented in this Letter should primarily be understood as providing a holographic or “external” description of the baby universe through the bulk entanglement of the quantum fields. Just below the MQ temperature, the bulk entanglement entropy of the baby universe equals the thermal entropy of the MQ Hamiltonian. Since this entropy is of $O(N)$, this suggests that the holographic encoding of the black hole interior and of the baby universe are alike in this model. In fact, for microcanonical versions of the TPS, both regimes of the wavefunction are connected by a smooth crossover \cite{Maldacena:2018lmt}. Only at very low temperatures $\mathcal{J}\beta \gtrsim O(\log N)$ does the entropy become non-extensive.   

The holographic description in this model (or in AS$^2$) suggests that there is a projection onto $\ket{\psi_{\Op}}$ occurring in the baby universe, analogous to the black hole final state \cite{Horowitz:2003he}. The state $\ket{\psi_{\Op}}$ may also be interpreted as a definite $\alpha$-state \cite{Marolf:2020xie}. Above, we have initiated a study of the relationship between the baby universe state $\ket{\psi_{\mathcal{O}}}$ and the SYK couplings at finite $N$. Extracting the physical implications of this projection in the semiclassical picture is the central open problem for elucidating the physical description of the baby universe for a fixed realization of the operator or couplings. We leave a more extensive study to \cite{usprep}.  \\

\textbf{Acknowledgments} — We thank Stefano Antonini, José Barbón, Alex Belin, Jeevan Chandra, Gabriele Di Ubaldo, Luca Iliesiu, Jonah Kudler-Flam, Henry Lin, Javier Magán, Henry Maxfield, Douglas Stanford, Pratik Rath, Moshe Rozali, Chris Waddell and Edward Witten for useful comments. This work is supported in part by the Leinweber Institute for Theoretical Physics. MS and BS acknowledge support from the U.S. Department of Energy through GeoFlow DE-SC0019380 and BS acknowledges support from the Heising-Simons Foundation through grant 2024-4849. AVL is supported by the National Science and Engineering Research Council of Canada (NSERC) and the Simons Foundation via a Simons Investigator Award.

\appendix

\section{Details on the semiclassical saddles}

\textbf{Black hole saddle} — The Euclidean trajectory of the $\ell$-particle has a conserved total energy in the inverted potential
\be 
\frac{1}{2}\dot{\ell}^2 - V(\ell) = -2\varepsilon\,,\qquad \varepsilon \equiv \dfrac{E}{N}\,.
\ee 
We define the turning length $\ell_{0}$ from the constraint $V(\ell_{0}) = 2\varepsilon$. Given this, it follows that
\be\label{eq:bounceconstrE} 
 \int_{-\infty}^{\ell_{0}}\dfrac{\text{d}\ell }{\sqrt{2(V(\ell)-2\varepsilon_\beta)}} = \frac{\beta}{2}\,,
\ee 
where $\beta$ is the dimensionless inverse temperature (SYK inverse temperature in units of $\alpha_S/\mathcal{J}$). Eq. \eqref{eq:bounceconstrE} determines the energy $E^{\rm d}_\beta = N \varepsilon_{\beta}$ as a function of $\beta$.

Let us be more explicit for $\Delta_H =1/2$ (free chiral fermion). We take $\kappa \equiv \Delta/N - \eta/2 > 0$ so there are no bound states. In this case, the bounce solution is \cite{Magan:2024aet}
\be\label{eq:sollengthint} 
\ell(u) =  2 \log \left(\frac{\sqrt{\kappa ^2 + 4 \varepsilon}}{2\varepsilon}\cos \left(u \sqrt{\varepsilon}\right) + \dfrac{\kappa}{2\varepsilon}\right) \,.
\ee
The particle reaches infinity at $u = \beta/2$, and this fixes
\be 
\cos(\dfrac{\beta}{2}\sqrt{\varepsilon_\beta}) = - \dfrac{\kappa}{\sqrt{\kappa^2 + 4\varepsilon_\beta}}\,.
\ee 

To evaluate the Euclidean on-shell action, we add a counterterm that cancels the divergence as $\ell \to - \infty$
\be
I_{\rm ct} = - 4 N e^{-\ell_{\partial}/2} + N \kappa (\ell_{\partial} - 2 \log N) \, ,
\ee
with $\ell_{\partial}$ a cutoff to be taken to $-\infty$. The $\ell_{\partial}$-independent constant has been chosen to match standard conventions in the quantized length theory; see \cite{usprep} for more details. Adding the topological piece of the action for the disk gives
\be
\frac{I[\Sigma_{\rm d}]}{N} = - \beta \varepsilon_{\beta} + 2 \kappa \left( 1 - \log \frac{N \sqrt{\kappa^2 + 4 \varepsilon_{\beta}}}{4} \right) - \frac{S_0}{N} \, .
\ee
At low preparation temperatures $\beta \kappa \gg 1$, we can solve for $\varepsilon_{\beta} \approx \frac{4 \pi^2}{\beta^2} - \frac{32 \pi^2}{\beta^4 \kappa^2} + \dots$. The action is
\be
I[\Sigma_{\rm d}] = - \frac{4 \pi^2 N}{\beta} + 2 N \kappa \left( 1 - \log \frac{N \kappa}{4} \right) - S_0 \, .
\ee

\begin{figure}[h]
    \centering
    \includegraphics[width=.9\linewidth]{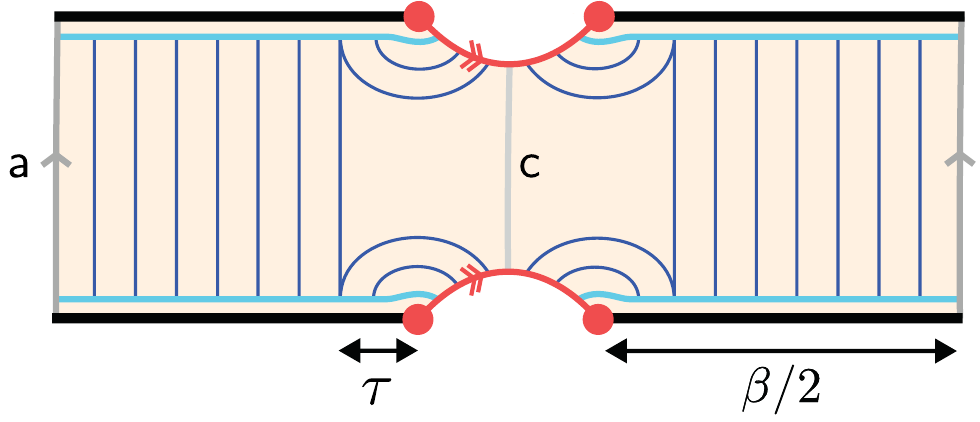}
    \caption{Uniformization of the handle. The blue lines correspond to minimal length geodesics between the boundary points.}
    \label{fig:handle}
\end{figure}

\textbf{Baby universe saddle} — The hyperbolic metric on $\Sigma_{\rm h}$ is uniformized on the disk by a subgroup of $\mathsf{PSL}(2,\mathbb{R})$ freely generated by two hyperbolic elements. Figure \ref{fig:handle} shows the corresponding fundamental domain of the subgroup, in coordinates on the disk adapted to the boundary particle’s trajectory.

Here, we will make some approximations that simplify the construction of this saddle; more details beyond this approximation will appear in \cite{usprep}. We work at low temperatures $\beta \eta^{1/(2(1-\Delta_H))} \gg 1$ and with a sufficiently heavy matter particle $\Delta/N \gg \eta^{1/(2(1-\Delta_H))}$. At such low temperatures, the Euclidean propagation in the ``empty'' MQ potential $V_0(\ell) = 2e^{-\ell} - {\eta}  e^{-\Delta_H \ell}$ prepares a configuration close to the ground state of the MQ Hamiltonian. This state resembles a finite temperature TFD of two decoupled SYK systems at the coupling-dependent inverse temperature $\beta_0 = \beta_0(\eta)$ given by \eqref{eq:invtemp}. 

Consequently, the remaining piece becomes approximately the two-boundary wormhole in $\text{Tr}(e^{-(\beta_0/2) H}\Op_\Delta)^2$, constructed in \cite{Stanford:2020wkf,Usatyuk:2024mzs}. We can read the size of the universe as the bottleneck modulus of the wormhole \cite{Usatyuk:2024mzs}, which at large $\beta_0 \Delta / N$ is given by
\be 
b(\Delta,\eta) \approx 2 \log \left(\dfrac{\Delta \beta_{0}  }{2\pi N} \right) \,.
\ee 

We compute the on-shell action again for $\Delta_H=1/2$. In the regime of parameters we are working with, we can approximate the Euclidean time $\tau$ at which the minimal geodesic length jumps by $\tau \approx 0$, meaning it is very close to the insertion of the operator $\Op_\Delta$. This is a good approximation when the universe is large, $\beta_0 \Delta /N \gg 1$. There is no propagation of the length in the presence of the operator, so the only contribution from $\Op_\Delta$ is the piece
\be
2 N \kappa \left( 1 - \log \frac{N \kappa}{4} \right) \, .
\ee
The rest is just the propagation of the $\ell$-particle in the empty MQ potential. At low temperatures $\beta \eta \gg 1$, this is dominated by the energy of the ground state; thus, the full action, including the topological term, is
\be
I[\Sigma_{\rm h}] \approx \beta E_0 + 2 N \kappa \left( 1 - \log \frac{N \kappa}{4} \right) + S_0 \, .
\ee
with $E_0$ given by \eqref{eq:energy}.\\

{\it State of the bulk fields.} In the same approximation as above, we want to write down the entangled state $\ket{\psi_\beta}$ prepared by the saddle $\Sigma_{\rm h}$ between the $\mathsf{a}$ and $\mathsf{c}$ components. Given an eigenstate $\ket{E_n}_{\mathsf{a}}$ of the MQ Hamiltonian in the gapped phase, we propagate it along the strip part of Fig. \ref{fig:handle} up to the insertion of the particle. This produces a term $\exp(-\beta E_n/2)$ in its wavefunction. The remaining part to reach $\mathsf{c}$ is to propagate it through the wormhole. This defines some state $\ket{\varphi_n}_{\mathsf{c}}\equiv e^{- \ell_\Delta H_{\rm bulk}/2}\ket{E_n}$, where $H_{\rm bulk}$ is the bulk Hamiltonian on the wormhole and $\ell_\Delta/2$ is the Euclidean time to the bottleneck of the wormhole. Doing this for each energy eigenstate in the gapped phase produces the TFD-like state \eqref{eq:TFDlikebulk} (we are assuming that the global charge of $\ket{E_n}$ satisfies \eqref{eq:Z4gauge}; otherwise, its component in the wavefunction vanishes).

\section{Other non baby universe TPSs}

One may attribute the very existence of the closed universe to the operator $\Op_\Delta$, which may play the role of the “observer” in recent discussions \cite{Abdalla:2025gzn,Harlow:2025pvj}. Indeed, the baby universe saddle exists even at $\beta \to \infty$, when the holographic description reduces to the ground state of the MQ Hamiltonian. Moreover, similar baby universes arise even in the absence of a nontrivial holographic Hilbert space \cite{Stanford:2020wkf,Usatyuk:2024mzs}. 

Here, we make an observation using our setup and consider a new TPS constructed from a different reference state
\be 
\ket{\Psi^P_\beta} \,\propto\, e^{-\frac{\beta}{2}H} \ket{P}\ket{P}\,,
\ee 
where $\ket{P}$ is a pure state in a single SYK model. The preparation of this new state is represented as\\[-.4cm]
\begin{center}
\includegraphics[width=0.46\linewidth]{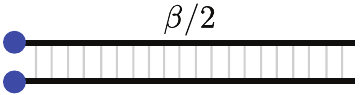}
\end{center}

We define the pure state to be the ground state of the $N/2$ Dirac fermions of a single SYK model
\be 
(\psi_{i} - \iw \psi_{N-i}) \ket{P} =0\,,\qquad \forall\; i=1,...,\frac{N}{2}\,.
\ee 
The state $\ket{P}$ can be modeled as a boundary condition associated with a high-energy end-of-the-world brane \cite{Kourkoulou:2017zaj}. 

The semiclassical dual to this new TPS $|\Psi^P_\beta\rangle$ transitions from a state of two single-sided black holes with end-of-the-world branes in the interior (as in \cite{Kourkoulou:2017zaj}) to a state of an eternal traversable wormhole at low temperatures. Notably, the low-temperature phase of $|\Psi^P_\beta\rangle$ does not contain a semiclassical baby universe.

This may reflect two features that characterize a good “observer”.  On the one hand, $\ket{P}\ket{P}$ is a state of Schmidt rank $1$, so it contains low entropy. This needs to be contrasted with $\ket{\Op_\Delta}$, which has a large Schmidt rank. The Hilbert space dimension computed in \cite{Abdalla:2025gzn, Harlow:2025pvj} is (partially in \cite{Abdalla:2025gzn}) associated with this entropy.

Furthermore, the states $|\Psi^\Delta_\beta\rangle$ and $|\Psi^P_\beta\rangle$ differ in that the former is quasi-random (see Fig. \ref{fig:c0}), whereas the latter is not. In particular, the MQ ground state component $\langle 0|(|P\rangle|P\rangle)$ admits a self-averaging semiclassical contribution, where the end-of-the-world brane connects the two pieces of the reference state. A closely related situation arises in the AS$^2$ model when the thin shell is replaced by a localized source of a heavy scalar primary (or an end-of-the-world brane), since such operators exhibit a semiclassical thermal one-point function.

It is worth noting that for $|\Psi_\beta^\Delta\rangle$ the heavy operator $\Op_\Delta$ lacks a semiclassical one-point function at the disk level in the JT description because of the emergent $\mathsf{PSL}(2,\mathbb{R})$ symmetry in the infrared. This may avoid the subtleties encountered in the higher-dimensional AS$^2$ model, where the absence of such a contribution is less obvious.

\bibliography{bibliography}
\bibliographystyle{ourbst}

\end{document}